\documentclass[12pt]{iopart}
\usepackage[dvips]{graphicx}
\usepackage{color}

\eqnobysec

\begin{document}


\title{Newton-Machian analysis of Neo-tychonian model of planetary motions}

\author{Luka Popov \vspace{2mm}}

\address{University of Zagreb, Department of Physics,
Bijeni\v cka cesta 32, Zagreb, Croatia}

\ead{lpopov@phy.hr}

\begin{abstract} \noindent
  The calculation of the trajectories in the Sun-Earth-Mars system will be
performed in two different models, both in the framework of Newtonian mechanics.
First model is well-known Copernican system, which assumes the Sun is at rest
and all the planets orbit around it. Second one is less-known model developed by
Tycho Brahe (1546-1601),
  according to which the Earth stands still, the Sun orbits around the Earth,
and other planets orbit around the Sun. The term ``Neo-tychonian system'' refers
to the assumption that orbits of distant masses around the Earth are
synchronized with the Sun's orbit. It is the aim of this paper to show the
kinematical and dynamical equivalence of these systems, under the assumption of
Mach's principle.
\end{abstract}



\submitto{\EJP}

\vspace{1.5ex}
\noindent{\footnotesize Accepted for publication: 16 Jan 2013}

 \maketitle

\setcounter{equation}{0}
\section{Introduction}\label{intro}

The discussion of motion of celestial bodies is one of the most interesting
episodes in the history of science. There are two diametrically opposite schools
of thought: one that assumes that the Sun stands still, and Earth and other
planets orbit around it; and another that assumes that the Earth stands still,
and Sun and other planets in some manner orbit around the Earth. The first
school of thought comes from Aristarchus (310-230 {\scshape bc}) and is generally
addressed as \emph{heliocentrism}, another from Ptolemy (90-168 {\scshape bc})
and is generally known as \emph{geocentrism}. Since Aristotle, the ultimate authority
in science for more than two millennia, accepted the geocentric assumption, it
became dominant viewpoint among scientists of the time. The turnover came with
Copernicus (so-called ``Copernican revolution'') who in his work \emph{De
Revolutionibus} proposed a hypothesis that the Sun stands in the middle of the
known Universe, and that Earth orbits around it, together with other planets.
Copernicus' system was merely better than Ptolemy's, because Copernicus assumed
the trajectories of the planets are perfect circles, and required the same
number of epicycles (sometimes even more) as Ptolemy's model \cite{koestler}.
The accuracy of Ptolomy's model is still a subject of vivid debates among
historians of science \cite{rawlins}.

The next episode in this controversy is Kepler's system with elliptical orbits
of planets around the Sun. That system did not require epicycles, it was precise
and elegant. It is therefore general view that Kepler's work finally settled the
question whether it is the Sun or the Earth that moves. But what is less known
is that Tycho Brahe, Kepler's tutor, developed a geostatic system that was just
as accurate and elegant as Kepler's: the Sun orbits around the Earth, and all
the other planets orbit around the Sun. The trajectories are ellipses, and all
the Kepler's laws are satisfied. In that moment of history, the Kepler's and
Brahe's models were completely equivalent and equally elegant, since neither of
them could explain the mechanism and reason why the orbits are the way they are.
It had to wait for Newton.

Sir Isaac Newton, as it is generally considered, gave ultimate explanation of
planetary motions that was in accord with Kepler's model, and excluded Brahe's
one. The laws of motions and the inverse square law of gravity could reproduce
all the observed data only with the assumption that the Sun (i.e. the center of
mass of the system, which can be very well approximated by the center of the
Sun) stands still, and all planets move around it. According to Newton's laws,
it is impossible for small Earth to keep the big Sun in its orbit: the
gravitational pull is just too weak. This argument is very strong, and it seemed
to settle the question for good.

But in the end of 19th century, the famous physicist and philosopher Ernst Mach
(1839-1916) came with the principle which states the equivalence of non-inertial
frames. Using the famous ``Newton's bucket'' argument, Mach argues that all
so-called pseudo-forces (forces which result from accelerated motion of the
reference frame) are in fact \emph{real} forces originating form the accelerated
motion of distant masses in the Universe, as observed by the observer in the
non-inertial frame. Some go even further, stating that ``every single physical
property and behavioral aspect of isolated systems is determined by the whole
Universe'' \cite{rosen}. According to Mach's principle, the Earth could be considered
as the ``pivot point'' of the Universe: the fact that the Universe is orbiting
around the Earth will create the exact same forces that we usually ascribe to
the motion of the Earth.

Mach's principle played a major role in the development of the Einstein's General
Theory of Relavity \cite{newburgh}, as well as other developments in gravitation theory,
and has inspired some interesting experiments \cite{licht}.
This principle still serves as a guideline for some physicists who
attempt to reformulate (``Machianize'') Newtonian dynamics \cite{hood, barbourN},
or try to construct new theories of mechanics \cite{assis}.
Some arguments and critiques against Mach's principle have also been raised
\cite{hartman}.
Since the time of it's original appearance \cite{mach1, mach2, mach3},
Mach's principle has been reformulated in numbers of different ways
\cite{rovelli, barbour}. For the purpose of this paper, we will only focus on
the one of the consequences of Mach's principle: that the inertial forces can be seen
as resulting from real interactions with distant matter in the Universe, as
was for example shown by A.~Zylbersztajn \cite{zylbersztajn}.

The only question remains: \emph{are these forces by themselves enough to
explain all translational motions that we observe from Earth,
and can they reproduce the Tycho Brahe's model?} The discussion in this paper
will show that the answer to this question is positive. In order to demonstrate
it, we will consider the Sun-Earth-Mars system.

The paper is organized as follows. In section \ref{twobody} an overview of
two-body problem in the central potential and of Kepler's problem is given. In
section \ref{helioc} the calculations of Earth's and Mars' trajectories are
performed in the heliocentric system, both analytically (by applying the results
from previous section) and numerically. In section \ref{geoc} the calculations
of Sun's and Mars' trajectories are performed in geocentric system, due to the
presence of pseudo-potential originating from the fact of accelerated motion of
the Universe. Finally, the conclusion of the analysis is given.

\section{Two-body problem in the central potential}\label{twobody}

\subsection{General overview}

We start with the overview of two body problem in Newtonian mechanics. Although
there are alternative and simpler ways to solve this problem \cite{hauser, gauthier},
we will follow the usual textbook approach \cite{landau, goldstein}.
The Lagrangian of the system reads
\begin{equation} \label{2bodyL}
L = \frac{1}{2} m_1 \dot{\mathbf{r}}_1^2 + \frac{1}{2} m_2 \dot{\mathbf{r}}_2^2
 - U(\left| \mathbf{r}_1 - \mathbf{r}_2 \right|) \,,
\end{equation}
where $U$ is potential energy that depends only on the magnitude of the
difference of radii vectors (so-called \emph{central potential}). We can easily
rewrite this equation in terms of relative position vector $\mathbf{r} \equiv
\mathbf{r}_1 - \mathbf{r}_2$, and let the origin be at the centre of mass, i.e.
$m_1 \mathbf{r}_1 + m_2 \mathbf{r}_2 \equiv 0$. Solution of these equations are
\begin{equation} \label{r1r2}
\mathbf{r}_1 = \frac{m_2}{m_1+m_2}  \mathbf{r} \,, \qquad
    \mathbf{r}_2 = - \frac{m_1}{m_1+m_2}  \mathbf{r} \,.
\end{equation}
The Lagrangian (\ref{2bodyL}) so becomes
\begin{equation} \label{redL}
L = \frac{1}{2} \mu \dot{\mathbf{r}}^2 - U(r) \,,
\end{equation}
where $r \equiv \left| \mathbf{r} \right|$ and $\mu$ is \emph{the reduced mass},
\begin{equation} \label{redmass}
\frac{1}{\mu} = \frac{1}{m_1} + \frac{1}{m_2}
\end{equation}
In that manner, the two-body problem is reduced to one-body problem of particle
with coordinate $\mathbf{r}$ and mass $\mu$ in the potential $U(r)$.

Using polar coordinates, the Lagrangian (\ref{redL}) can be written as:
\begin{equation}
L = \frac{1}{2} \mu (\dot{r}^2 + r^2 \dot{\phi}^2) - U(r)
\end{equation}
One can immediately notice that variable $\phi$ is cyclic (it does not appear in
the Lagrangian explicitly). Consequence of that fact is momentum conservation
law, since $(\partial / \partial t) (\partial L / \partial \dot{\phi}) =
\partial L / \partial \phi = 0$. Therefore,
\begin{equation} \label{momentum}
\ell \equiv \frac{\partial L}{\partial \dot{\phi}} = \mu r^2 \dot{\phi} =
\mathrm{const.}
\end{equation}
is the integral of motion.

In order to find a solution for the trajectory of a particle, it is not
necessary to explicitly write down the Euler-Lagrange equations. Instead, one
can use the energy conservation law,
\begin{equation} \label{zoe}
E = \frac{1}{2} \mu (\dot{r}^2 + r^2 \dot{\phi}^2) + U(r)
    = \frac{1}{2} \mu \dot{r}^2 + \frac{\ell^2}{2 \mu r^2} + U(r)
\end{equation}
Straightforward integration of (\ref{zoe}) gives equation for the trajectory,
\begin{equation} \label{gentraj}
\phi(r) = \int \frac{ \ell \,\, \mathrm{d} r / r^2 }
    { \sqrt{ 2\mu \left[ E-U(r) \right] - \ell^2/r^2 } }
\end{equation}

\subsection{Kepler's problem}

Let us now consider the particle in the potential
\begin{equation} \label{kepler}
U(r) = - \frac{k}{r} \,,
\end{equation}
generally known as \emph{Kepler's problem}. Since our primary interest is in the
planetary motions under the influence of gravity, we will take $k>0$. Integration
of eq. (\ref{gentraj}) for that potential gives:
\begin{equation} \label{elipsa}
\frac{p}{r} = 1 + e \cos \phi \,,
\end{equation}
where $2p$ is called \emph{lactus rectum} of the orbit, and $e$ is
\emph{eccentricity}. These quantities are given by
\begin{equation} \label{pe}
p = \frac{\ell^2}{\mu k} \,, \qquad e=\sqrt{ 1 + \frac{2 E \ell^2 }{ \mu k^2 }
}
\end{equation}
Expression (\ref{elipsa}) is the equation of a conic section with one focus in
the origin. For $E<0$ and $e<1$ the orbit is an ellipse.

One can also determine minimal and maximal distances from the source of the
potential, called \emph{perihelion} and \emph{aphelion}, respectively:
\begin{equation} \label{rminmax}
r_{min} = \frac{p}{1+e} \,, \qquad
    r_{max} = \frac{p}{1-e} \,.
\end{equation}
These parameters can be directly observed, and often are used to test a model or
a theory regarding planetary motions.

\section{Earth and Mars in heliocentric perspective}\label{helioc}

According to Newton's law of gravity, the force between two massive objects
reads:
\begin{equation}
\mathbf{F} = - \frac{G m_1 m_2}{ \left| \mathbf{r}_1 - \mathbf{r}_2 \right|^3 }
 ( \mathbf{r}_1 - \mathbf{r}_2 ) \,.
\end{equation}
Which leads to a potential ($\mathbf{F}=-\nabla U$)
\begin{equation}
U(\left| \mathbf{r}_1 - \mathbf{r}_2 \right|) =
    - \frac{G m_1 m_2}{ \left| \mathbf{r}_1 - \mathbf{r}_2 \right| } \,.
\end{equation}
This is obviously Kepler's potential (\ref{kepler}) with $k = G m_1 m_2$, where
$G$ is Newton's gravitational constant.

Since the Sun is more than 5 orders of magnitude more massive than Earth and
Mars, we will in all future analysis use the approximation
\begin{equation}
\mu \approx m_i \,,
\end{equation}
where $m_i$ is mass of the observed planet. For the same reason, gravitational
interaction between Earth and Mars can be neglected, since it is negligible
compared with the interaction between Earth/Mars and the Sun.

\begin{figure}[t]
  \centering
  \includegraphics[scale=0.7]{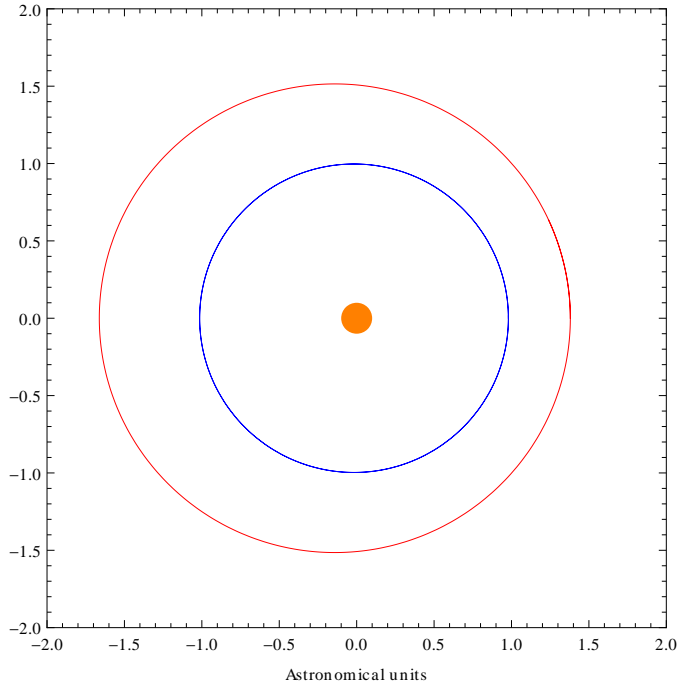}
  \caption{\label{Fhelio} Trajectories of Earth and Mars in heliocentric system
over the period of 2 years. Blue and red lines represent Earth's and Mars'
orbits, respectively (color online).}
\end{figure}

Using these assumptions, we can write down corresponding Lagrangians,
\begin{eqnarray}
L_{ES} &=& \frac{1}{2} m_E \dot{\mathbf{r}}_{ES}^2 + \frac{G m_E M_S}{r_{ES}}
\,,
\nonumber \\
L_{MS} &=& \frac{1}{2} m_M \dot{\mathbf{r}}_{MS}^2 + \frac{G m_M M_S}{r_{MS}}
\,,
\label{helioLang}
\end{eqnarray}
where $m_E$ and $m_M$ are masses of Earth and Mars, respectively. Subscripts
$ES$ ($MS$) correspond to the motion of Earth (Mars) with respect to the Sun.
These trajectories can be calculated using the exact solution (\ref{elipsa})
with appropriate strength constants $k$ and initial conditions which determine
$E$ and $\ell$. Another way is to solve the Euler-Lagrange equations
numerically, using astronomical parameters \cite{handbook} (e.g. aphelion and
perihelion of Earth/Mars) to choose the inital conditions that fit the observed data.
The former has been done using \emph{Wolfram Mathematica} package.
The result is shown on Fig.\ \ref{Fhelio}.

For latter comparison, one could write out the expressions for the $e$ and
$p$ parameters for the Earth. Putting the expressions for energy (\ref{zoe}) and
momentum (\ref{momentum}) into Equations (\ref{pe}) it is straightforward to obtain
\begin{eqnarray} \label{peexact}
p &=& \frac{ \dot{\phi}^2 r^4 }{ G M_S } \,, \nonumber \\ \nonumber & & \\
e &=& \sqrt { 1 - \frac{2 G M_S \dot{\phi}^2 r^3 - \dot{r}^2 \dot{\phi}^2 r^4
        - \dot{\phi}^4 r^6}{ G^2 M_S^2}  } \,,
\end{eqnarray}
where $\dot{\phi}$, $\dot{r}$ and $r$ are angular velocity, radial velocity and
distance respectively, taken in the same moment of time (e.g. in $t=0$).

\begin{figure}[t]
  \centering
  \includegraphics[scale=0.7]{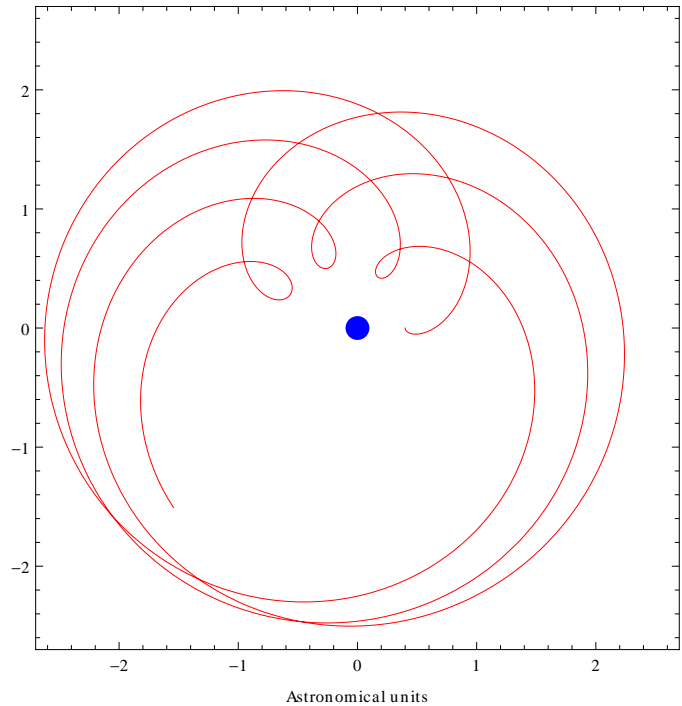}
  \caption{\label{FtychoM} Trajectory of the Mars as seen from the Earth over
the period of 7 years. Calculation of this trajectory is done numerically in the
heliocentric system.}
\end{figure}

Fig.\ \ref{FtychoM} displays motion of the Mars as viewed from the Earth, gained
by trivial coordinate transformation
\begin{equation} \label{coord}
\mathbf{r}_{ME}(t) = -\mathbf{r}_{ES}(t) + \mathbf{r}_{MS}(t) \,,
\end{equation}
where $\mathbf{r}_{ES}(t)$ and $\mathbf{r}_{MS}(t)$ are solutions of
Euler-Lagrange equations for the Lagrangians (\ref{helioLang}). Equation (\ref{coord})
is just the mathematical expression of the Tycho Brahe's claim. Retrograde motion
of the Mars can be useful in the attempt to understand and determine orbital
parameters, as was qualitative and quantitative shown by B.~Thompson \cite{thompson}.

The acceleration that Earth experiences due to the gravitational force of the
Sun is usually referred as \emph{centripetal acceleration} and is given by
\begin{equation} \label{acp}
\mathbf{a}_{cp} = \frac{\mathbf{F}_{cp}}{m_E} = - \frac{G M_S}{r_{ES}^2}
    \hat{\mathbf{r}}_{ES} \,,
\end{equation}
where $\hat{\mathbf{r}}$ is the unit vector in the direction of vector
$\mathbf{r}$, $\mathbf{r}_{ES}(t)$ is radius vector describing motion of Earth
with respect to the Sun, and $\mathbf{F}_{cp}$ is centripetal force,
i.e.\ the force that causes the motion.

\section{Sun and Mars in geocentric perspective} \label{geoc}

\subsection{The pseudo-potential}

From the heliocentric perspective, the fact that the Earth moves around the Sun
results with centrifugal pseudo-force, observed only by the observer on the
Earth. But if we apply Mach's principle to the geocentric viewpoint, one is obliged
to speak about the \emph{real} forces resulting from the fact that the Universe
as a whole moves around the observer sitting on the stationary Earth. Although these
forces will further be considered as the real forces, we well keep the usual
terminology and call them pseudo-forces, for the sake of convenience. Our focus
here will be on the annual orbits, not on diurnal rotation which requires some
additional physical assumptions \cite{assis} \cite{veto} that are beyond the
scope of this paper.

The Universe is regarded as an $(N+1)$-particle system ($N$ celestial bodies plus
planet Earth). From the point of a stationary Earth, one can write down the
Lagrangian that describes the motions of celestial bodies:
\begin{equation} \label{LUniv}
L = \frac{1}{2} \sum_{i=1}^{N} m_i \dot{\mathbf{r}}_i^2 -
    \frac{1}{2} \sum_{i=1}^{N}\frac{G m_i m_j}{r_{ij}} -
    \sum_{i=1}^{N} \frac{G m_E m_i}{r_i} - U_{ps} \,,
\end{equation}
where $r_{ij} \equiv |\mathbf{r}_i - \mathbf{r}_j|$, $U_{ps}$ stands for
pseudo-potential, satisfying $\mathbf{F}_{ps} = - \nabla U_{ps}$.
$\mathbf{F}_{ps}$ is the pseudo-force given by
\begin{equation} \label{Fpsgen}
\mathbf{F}_{ps} = - m \sum_{i=1}^{N} \mathbf{a}_{cp,i}  \,\,,
\end{equation}
where $\mathbf{a}_{cp,i}$ is centripetal acceleration for given celestial body
(with respect to the Earth) and $m$ is a mass of the object that is subjected to
this force. It's easy to notice that the dominant contribution in these sums
comes from the Sun. The close objects (planets, moons, etc) are much less
massive than the Sun, and massive object are much further away. The same
approximation is implicitly used in section \ref{helioc}.

In the Machian picture, the centripetal acceleration is a mere relative
quantity, describing the rate of change of relative velocity. Therefore,
centripetal acceleration of the Sun with respect to Earth is given by
Equation (\ref{acp}), with $\mathbf{r}_{ES}=-\mathbf{r}_{SE}$. All that considered,
Equation (\ref{Fpsgen}) becomes
\begin{equation}
\mathbf{F}_{ps} = - \frac{G m M_S}{r_{SE}^2} \hat{\mathbf{r}}_{SE} \,,
\end{equation}
where $\mathbf{r}_{SE}(t)$ describes the motion of the Sun around the Earth,
and $m$ is the mass of the body under consideration.

We can now finally write down the pseudo-potential which influences every body
observed by still observer on Earth:
\begin{equation} \label{Ups}
U_{ps} (\mathbf{r}) = \frac{G m M_S}{r_{SE}^2} \hat{\mathbf{r}}_{SE} \cdot
\mathbf{r} \,,
\end{equation}
where $\mathbf{r}(t)$ describes motion of particle of mass $m$ with respect to
the Earth. Notice that this is not a central potential.

\subsection{Sun in Earth's pseudo-potential}

In order to determine Sun's orbit in Earth's pseudo-potential, one needs to take
dominant contributions of the Lagrangian (\ref{LUniv}), as was explained earlier.
Taking into account the expression for pseudo-potential given in Equation (\ref{Ups}),
one ends up with
\begin{equation} \label{LSE}
L_{SE} = \frac{1}{2} M_S \dot{\mathbf{r}}_{SE}^2 - \frac{ G M_S^2 }{ r_{SE} }
\,.
\end{equation}
This Lagrangian has the exact same form as the reduced Lagrangian (\ref{redL}).
That means that we can immediately determine the orbit by means of Equations (\ref{pe})
by substituting $\mu = M_S$ and $k = G M_S^2$. This leads to the
following result (subscript $SE$ will be omitted):
\begin{eqnarray}
p &=& \frac{ \dot{\phi}^2 r^4 }{ G M_S } \,, \nonumber \\ \nonumber & & \\
e &=& \sqrt { 1 - \frac{2 G M_S \dot{\phi}^2 r^3 - \dot{r}^2 \dot{\phi}^2 r^4
        - \dot{\phi}^4 r^6}{ G^2 M_S^2}  } \,,
\end{eqnarray}
which is the exact equivalent as the previous result given in Equations (\ref{peexact}), since
$\dot{\phi}$, $\dot{r}$ and $r$ are relative quantities, by definition
equivalent in both models. We can therefore conclude that the Sun's orbit in the
Earth's pseudo-potential is equivalent to that observed from the Earth in the
heliocentric system.

It remains to show the same thing for Mars' orbit.

\subsection{Mars in Earth's pseudo-potential}

In the similar way as before, we take the dominant contributions of Lagrangian
(\ref{LUniv}) together with Equation (\ref{Ups}) and form the following Lagrangian
\begin{equation} \label{LME}
L_{ME} = \frac{1}{2} m_M \dot{\mathbf{r}}_{ME}^2 +
 \frac{G m_M M_S}{| \mathbf{r}_{ME} - \mathbf{r}_{SE} |} -
 \frac{G m_M M_S}{r_{SE}^2} \hat{\mathbf{r}}_{SE} \cdot \mathbf{r}_{ME} \,,
\end{equation}
where subscript $ME$ refers to the motion of Mars with respect to the Earth, and
$\mathbf{r}_{SE}(t)$ is the solution of Euler-Lagrange equations for the Lagrangian
({\ref{LSE}).

\begin{figure}[ht!]
  \centering
  \includegraphics[scale=0.7]{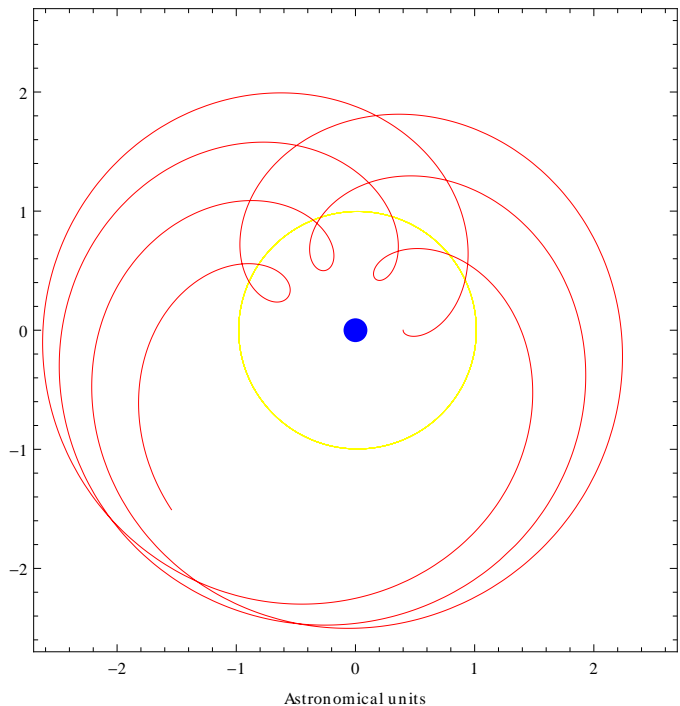}
  \caption{\label{Fgeoc} Trajectories of the Sun (yellow) and the Mars (red)
moving in Earth's pseudo-potential over the period of 7 years (color online).
Calculation of this trajectory is performed numerically in the geocentric system.}
\end{figure}

The Euler-Lagrange equations for $\mathbf{r}_{ME}(t)$ using
Lagrangian (\ref{LME}) are too complicated to be
solved analytically, but they can easily be solved numerically. The numerical
solutions for the equations of motion for both the Sun and Mars are displayed in
Fig.~\ref{Fgeoc}. The equivalence of trajectories gained in two different ways
is obvious, justifying the model proposed by Tycho Brahe.

\section{Conclusion}

The analysis of planetary motions has been performed in the Newtonian framework with
the assumption of Mach's principle. The kinematical equivalence of the Copernican
(heliocentric) and the Neo-tychonian (geocentric) systems is shown to be a
consequence of the presence of pseudo-potential (\ref{Ups}) in the geocentric
system, which, according to Mach, must be regarded as the real potential
originating from the fact of the simultaneous acceleration of the Universe. This
analysis can be done on any other celestial body observed from the Earth. Since
Sun and Mars are chosen arbitrarily, and there is nothing special about Mars, one
can expect to come up with the same general conclusion.

There is another interesting remark that follows from this analysis. If one
could put the whole Universe in accelerated motion around the Earth, the
pseudo-potential corresponding to pseudo-force (\ref{Fpsgen}) will immediately
be generated. That same pseudo-potential then causes the Universe to stay in
that very state of motion, without any need of exterior forces acting on it.

{\ack This work is supported by Ministry of Science, Sports and Technology under contract
number 119-0982930-1016.}

\end{document}